\documentclass[12pt, pra, aps]{revtex4}

\usepackage{chemformula} 
\usepackage[T1]{fontenc}
\usepackage{upgreek}
\usepackage{comment}
\usepackage{multirow}
\usepackage{nameref,zref-xr}
\zxrsetup{toltxlabel}
\usepackage{hyperref}
\zexternaldocument*[SI-]{SI-HClonGold-rev} 
\setcitestyle{super}

\begin{document}
\title{Dissociative Electron Attachment on Metal Surfaces: The Case of \ch{HCl-} on Au(111)}
\author{Robin E. Moorby,$^{a,b}$ Valentina Parravicini,$^a$ Thomas-C. Jagau,$^{a}$ and Maristella Alessio$^{\ast a}$\\
{\small $^a$Department of Chemistry, KU Leuven, Celestijnenlaan 200F, B-3001 Leuven, Belgium}\\
{\small $^b$Department of Chemistry, University of Durham, South Road, Durham, DH1 3LE, United Kingdom}\\
$^{\ast}$ Corresponding author: maristella.alessio@kuleuven.be}
\begin{abstract}
The transfer of charges, including electrons and holes, is a key step in heterogeneous catalysis, taking part in the reduction and oxidation of adsorbate species on catalyst surfaces. 
In plasmonic catalysis, electrons can transfer from photo-excited metal nanoparticles to molecular adsorbates, forming transient negative ions that can
easily undergo reactions such as dissociation, desorption, or other chemical transformations. 
However, \textit{ab initio} characterization of these anionic states has proven challenging, and little is known about the topology of their potential energy surfaces. 
In this work, we investigate the dissociative adsorption of HCl on Au(111) as a representative catalytic process with relatively low reaction probabilities, which could potentially be enhanced by electron transfer from photo-excited gold nanoparticles to HCl.
We employ projection-based density embedding that combines the equation-of-motion electron-attachment coupled-cluster singles and doubles (EOM-EA-CCSD) method with density functional theory (DFT), and build dissociation curves of HCl$^-$ on Au(111) along the H--Cl bond distance.
The HCl anion in the gas phase is unbound at equilibrium distances and only becomes bound as the bond stretches.
However, our results show that, upon adsorption on Au(111), HCl$^-$ remains a stable, bound anion at all bond lengths due to charge delocalization to the metal.
Forming bound anions is easier, and dissociation of HCl$^-$ on Au(111) is further facilitated, with its dissociation energy reduced by 0.61 eV compared to its neutral counterpart on Au(111), and by 1.16 eV relative to HCl.
These results underscore the efficacy of embedded EOM-CCSD methods in addressing surface science challenges and highlight the potential of plasmonic catalysis proceeding via bound, rather than transient, anionic states.
\end{abstract}

\maketitle

\section{Introduction}

Charge transfer between solid surfaces and adsorbate species is crucial in a wide range of chemical processes, including electronics and optoelectronics,\cite{Otero:CTOLEDS:2017,Dong:CTOLEDS:2021} redox cycles in solid oxide fuel cells,\cite{Goodwin:CTSOFC:2009,Bak:CTSOFC:2019} and photocatalysis on semiconductors\cite{Kudo:Semiconductors:2009,Chen:Semiconductors:2021} or plasmonic metal nanoparticles.\cite{Linic:PlasmonicReview:2011,Brongersma:PlasmonicReview:2015,Aslam:Plasmonic:2018,Cortes:PlasmonReview:2020,MartinezCarter:PlasmonReview:2021}
In plasmonic catalysis, sunlight is harnessed to generate surface plasmon resonances, which catalyze chemical reactions through both thermal and non-thermal pathways. 
Unlike metal surfaces that require high laser intensities to be photo-excited and initiate chemistry, plasmonic metal nanoparticles have shown catalytic activity using visible light at room temperature.\cite{PhillipLinic:DeltaDFTO2:2011,MukherjeeCarter:PlasmonExptTheory:2012,Christopher:O2Ag:2012,Linic:O2:2013}
Once generated, the plasmon resonance can decay into energetic electrons, which either transfer to the adsorbate or dissipate their energy as heat into the environment. 
In the case of electron transfer, charged or excited states in the adsorbate can form, triggering chemical reactions by reducing the energy barrier along their potential energy surfaces.

This electron transfer is most often understood as an electronic excitation within the adsorbate-metal complex, i.e. from the metal Fermi level to an antibonding orbital in the adsorbate, resulting in excited states of charge-transfer character.\cite{Linic:PlasmonicReview:2011,Aslam:Plasmonic:2018}
Consequently, the adsorbate reactant progresses along its excited-state potential energy surface, experiencing a lower energy barrier, or decays to its ground state with additional vibrational energy, thus also facilitating the reaction. 
Following this excited state-mediated mechanism, density functional theory (DFT) with periodic boundary conditions (pbc) has initially been employed to construct ground and excited-state potential energy surfaces for molecules on metal surfaces.\cite{Gavnholt:DeltaDFT:2008,Olsen:DeltaDFT:2009,PhillipLinic:DeltaDFTO2:2011}
Additionally, the availability of low-lying, easy to populate antibonding orbitals on the adsorbate has been assessed from the electron density of states of the adsorbate-metal system using periodic DFT.\cite{PhillipLinic:DeltaDFTO2:2011,MukherjeeCarter:PlasmonExptTheory:2012}
In some cases, the transient nature of the anionic reactant molecule in the gas phase has been explored using CASPT2.\cite{MukherjeeCarter:PlasmonExptTheory:2012}
Furthermore, embedded correlated wavefunction calculations, which combine complete active space perturbation theory (CASPT2)\cite{Roos:92:CASPT2} with periodic DFT, have been performed, making it possible to handle 
charge-transfer excited states in larger %molecule-metal surface 
systems with greater accuracy.\cite{MukherjeeCarter:PlasmonExptTheory:2012,LibischCarter:H2:2013,Martinez:N2:2016,Bao:NH3:2019}

This charge-transfer excited state is commonly referred to as a transient negative ion.\cite{Linic:PlasmonicReview:2011,MukherjeeCarter:PlasmonExptTheory:2012,Brongersma:PlasmonicReview:2015} 
However, the associated reaction mechanism does not entail electron extraction from the metal, neither direct electron attachment to the adsorbate, thus the adsorbate-metal complex is not a negatively charged state. 
Therefore, \textit{ab initio} characterization of anionic states on surfaces has remained elusive, and their transient nature has yet to be conclusively demonstrated, as their properties are not expected to simply mirror those of their anionic counterparts in the gas phase. 
Moreover, experimental observation of charge-transfer states has also been challenging due to the difficulty of performing \textit{in operando} measurements of the electronic states and reaction kinetics of gas-metal surface reactions.\cite{Jianyu:CT:2020}

The electron attachment from the metal to an antibonding orbital of the adsorbate can lead to either bound (stable) or unbound (transient) anions. 
When the electron affinity of the adsorbate-metal system is positive, the electron attachment is permanent and the anion is bound.
In the molecular domain, the state of the art of treating bound anions is the equation-of-motion electron-attachment coupled-cluster (EOM-EA-CC) method.\cite{Nooijen:EOMEA:1995,posthf} 
However, already at the singles and doubles level of truncation (EOM-CCSD), its computational cost scales as $N^6$ with the number of basis functions, thereby hampering the applicability of this method for large molecules that require extended basis sets, which is often the case when addressing anions.
For periodic systems, wavefunction-based techniques have made remarkable progress in recent years and several studies have used EOM-CCSD to compute both neutral\cite{Wang:EOMCCpbc:2020,Gallo:EOMCCpbc:2021,Faruk:EOMCCpbc:2022} and charged\cite{McClain:EOMCCpbc:2016,McClain:EOMCCpbc:2017,Gao:EOMCCpbc:2020,Pulkin:EOMCCpbc:2020} excitation energies in a broad range of materials, including semiconductors, alkaline or transition-metal oxides, and two-dimensional quantum dots.
However, the applicability of periodic EOM-CCSD is significantly influenced by the number of atoms in the unit cell and size of the virtual orbital space. 
Performing these calculations thus remains a challenging task, particularly when dealing with extended systems, such as catalysts containing several hundreds of atoms in the supercell, as well as bulk metals or their nanostructures. 

For local phenomena, such as molecules on surfaces, quantum embedding theories
\cite{Wesolowski:QE:1993,Govidin:QE-pbc:1998,Knizia:DME:2012,Manby:QE-proj:2012,Alessio:CH4:2018,Sebastien:QE-proj:2019,Leighton:QE:2020} can be employed more conveniently.
%are more conveniently employed. 
Among these approaches, a groundbreaking contribution was made by introducing a density-based embedding technique that combines 
cluster models described by wavefunction theories with periodic DFT environments.\cite{Govidin:QE-pbc:1998,Huang:QE-pbc:2011,Libisch:QE-pbc:2014}
In this work, we employ the more recent projection-based version of density embedding,\cite{Manby:QE-proj:2012,Sebastian:QE-proj-gradients:2019,Sebastien:QE-proj:2019} 
which has been extensively used to explore ground and excited states\cite{Bennie:QM-proj-exc:2017} of both isolated molecules and periodic systems.\cite{Chulhai:QE-proj-pbc:2018}
Additionally, it has been extended to open-shell species\cite{Goodpaster:QE-proj-OS:2012} and recently been employed to describe spin states in transition-metal molecular magnets.\cite{Alessio:QE-proj-OS:2024}
Other advancements have shown the efficacy of projection-based embedded EOM-CCSD in accurately describing core and valence ionizations, valence excitations, and Rydberg states.\cite{Parravicini:QE-proj-EOM-CC:2021} 
Embedded EOM-EA-CCSD has also been used to treat anions and compute electron affinities.\cite{Parravicini:QE-proj-EOM-CC:2021} 
However, when electron affinities are small, the reliability of the results must be verified against EOM-CCSD for the full system.\cite{Parravicini:QE-proj-EOM-CC:2021} 
The recent availability of embedded EOM-EA-CCSD,\cite{Parravicini:QE-proj-EOM-CC:2021} combined with its applicability to molecules on metal surfaces\cite{Welborn:QE-proj-COonMe:2018}---two features not previously combined---has enabled this work.

In this work, we employ embedded CCSD and embedded EOM-EA-CCSD to investigate the dissociative adsorption of both neutral and anionic HCl on the Au(111) surface, respectively. 
Our interest in HCl/Au(111) is twofold.
First, the gas-phase barrier to HCl dissociation is too high to be overcome by thermal activation due to a deep well.
Additionally, experimental evidence suggests that HCl does not easily dissociate on Au(111) surfaces either.\cite{Shirhatti:Exp:2016}
A thorough screening of various DFT functionals has revealed a significant dependency of the dissociation barrier of HCl on Au(111) on the chosen functional,
 with DFT generally overestimating experimentally derived reaction probabilities.\cite{Fuchsel:PBE:2016,Fuchsel:PBE:2019,Gerrit:DFT:2020}
In this work, we go beyond DFT and employ embedded EOM-EA-CCSD to explore whether transfer of energetic electrons from the gold surface plasmon resonance to the $\upsigma ^{\ast}$ antibonding orbital of HCl, forming HCl$^{-}$/Au(111), can enhance HCl dissociation. 
Second, in the gas phase, the HCl molecule does not support a bound anionic state at its equilibrium structure.\cite{Nesbet:VirtualState:1977,Cizek:VirtualHCl:1999,Hotop:VirtualState:2003,Fedor:VirtualHCl:2010,Moorby:HCl:2024}  
However, upon bond stretching, the excess electron experiences weak stabilization by correlation, resulting in a bound HCl$^-$ state.\cite{Moorby:HCl:2024}  
Given that a chemical environment such as Au(111) can significantly alter the properties of molecular anions, our goal is to determine whether the dissociation of HCl$^{-}$ on Au(111) occurs via a bound or unbound anion. 

We demonstrate that the adsorption of HCl on Au(111) leads to sufficient delocalization of the extra electron in HCl$^-$ to the metal, such that the anionic state is bound at all bond lengths. 
We show how this process can favor the dissociation of HCl relative to its neutral form, both on the surface and in the gas phase.
Thus, by directly investigating the potential energy surface of the anionic adsorbate-metal complex, we go a step further than previous studies that focused on the excited state-mediated mechanism, and provide insights into additional plasmon-induced reaction pathways involving bound anionic states. 

\section{Computational details}
\label{sec:comput_details} 

\subsection{Periodic DFT calculations}

\begin{figure}[h!]
    \centering
    \includegraphics[width=8cm]{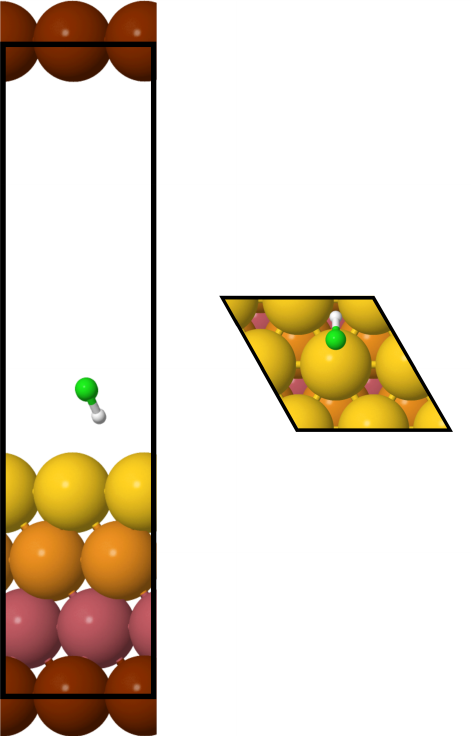}
    \caption{Side (left) and top (right) view of the $2\times2$ supercell used, including four gold layers and one HCl molecule per cell. Color code: H -- white, Cl -- green, Au (first to fourth layer) -- yellow to brown.}
    \label{fig:HClonAu}
\end{figure}

We began by performing periodic DFT structure optimizations of the HCl/Au(111) system. 
The Au(111) surface model was constructed by cutting the DFT optimized bulk cell along the (111) plane. 
For the gold bulk, we used a primitive face-centred cubic (\emph{fcc}) unit cell and for the Au(111) surface, we employed a $2\times2$ supercell formed by four layers of gold atoms. 
The periodic HCl/Au(111) adsorption system is shown in Figure \ref{fig:HClonAu}, consisting of one HCl molecule per surface cell. 
The lattice vector and the positions of the atoms in the bottom gold layer were kept frozen during structure optimization to simulate the bulk. 
The bulk and surface structures were optimized using the PBE\cite{PBE} functional, a plane-wave basis set, and projector-augmented wave (PAW) pseudopotentials\cite{Bloechl:PAW:1994,Kresse:PAW:1996} for the core electrons, as implemented in the Vienna ab initio simulation package (VASP version 5.4.4).\cite{VASP1,VASP2,VASP3,VASP4} 
In addition, at the PBE equilibrium structure of HCl/Au(111), we investigated the effect of van der Waals interactions by performing single-point calculations using the D3 correction scheme.\cite{Grimme:D3:2010}

In plane-wave DFT calculations, the quality of the basis set is determined by the plane-wave energy cut-off and volume of the unit cell.
The latter reduces, for surface models, to optimizing the vacuum height in the $z$ direction, to prevent artificial Coulomb interactions with the periodically repeated surface images.
For all calculations, an energy cut-off of 400 eV was employed. 
A vacuum space of 16 {\AA} was included in the $z$-direction of the periodic adsorption model. 
Likewise, to model the HCl molecule in the gas phase, we used a cubic cell of 25 {\AA} to ensure enough distance between the periodically repeated HCl molecules. 
Furthermore, achieving convergence in periodic DFT calculations necessitates ensuring that an adequate number of $k$-points are used to sample the first Brillouin zone.
For the gold bulk, we used an equally-spaced mesh of 17 $k$-points in each direction of the reciprocal space, centered at the $\Gamma$ point.
For all surface models, a $\Gamma$-centered $11\times11\times1$ $k$-point mesh was used.
For HCl, DFT calculations were performed using a $1\times1\times1$ $k$-point mesh centered at the $\Gamma$ point. 
For faster convergence with respect to the number of $k$-points, calculations involving metallic gold atoms employed the first-order Methfessel-Paxton smearing method\cite{Methfessel:Smearing:1989} with a smearing width of 0.2 eV. 
On the contrary, calculations on the HCl molecule used Gaussian-like smearing functions with a width of 0.01 eV. 
Additionally, harmonic vibrational frequencies of HCl and HCl on Au(111) were numerically computed using a centered finite difference scheme with a step width of 0.015 {\AA}.
All relevant Cartesian coordinates are provided in the Supplementary Information (SI).
Convergence tests with respect to the energy cut-off, $k$-point mesh, and vacuum size are reported in Figures S1-S4.

\subsection{Embedded EOM-CCSD calculations}

To build the dissociation curves of both neutral and anionic HCl on Au(111), we applied embedded EOM-EA-CCSD to a cluster model representing HCl/Au(111), and computed a manifold of electron-attached states and their corresponding electron-attachment energies at different H--Cl bond lengths. 
Within this EOM-EA-CCSD approach, the anionic states are obtained by applying electron-attachment operators to the neutral ground CCSD reference state.
As the reaction coordinate for HCl dissociation on Au(111), we considered the physisorbed HCl structure at equilibrium and systematically varied the position of the hydrogen atom along the H--Cl bond while keeping the remaining atoms frozen.

\begin{figure*}[h!]
    \centering
    \includegraphics[width=16cm]{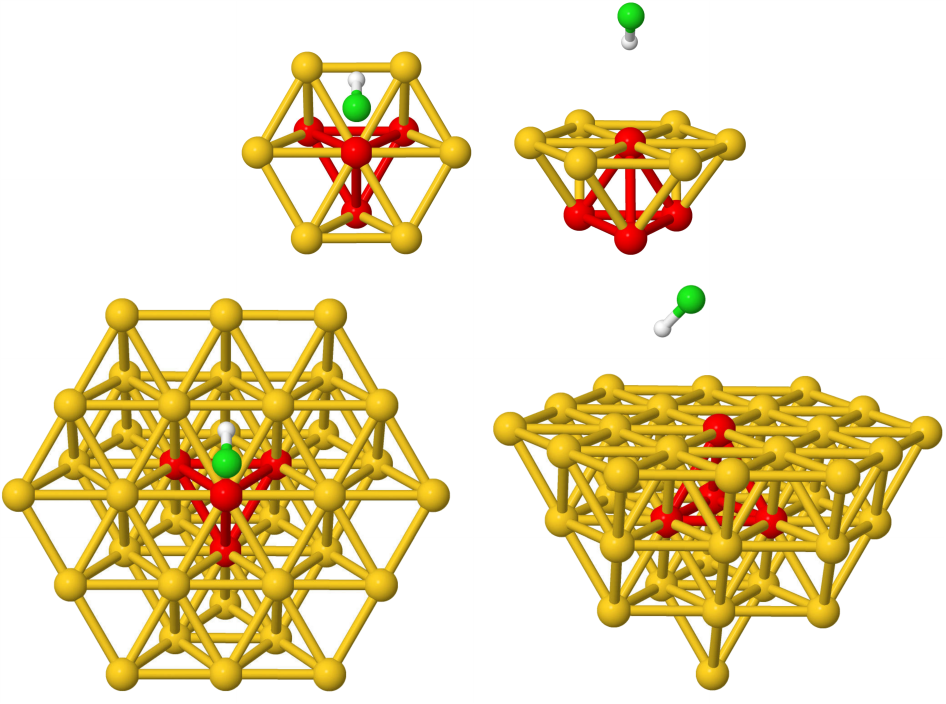}
    \caption{Top and side views of HCl/Au\(_{10}\) (top) and HCl/Au\(_{38}\) (bottom) cluster models. Color code: H -- white, Cl -- green, Au atoms belonging to the low-level DFT fragment -- yellow, Au atoms belonging to the high-level EOM-EA-CCSD fragment -- red}
    \label{fig:Au-clustermodels}
\end{figure*}

The adsorption cluster model comprises one HCl molecule positioned centrally onto a cluster of the Au(111) surface, which is cut out from the periodic PBE equilibrium structure.
We adopted two cluster models of the surface of different size: a two-layer 10-atom gold cluster (Au\(_{10}\)) and a four-layer 38-atom gold cluster (Au\(_{38}\)), as illustrated in Figure \ref{fig:Au-clustermodels}.
The Au\(_{38}\) cluster represents a natural extension of the \ch{Au10} cluster, including an additional layer of gold atoms. 
These clusters were designed to have an even number of gold atoms, thus maintaining a spin multiplicity of one. 
Similar clusters have been previously used in CCSD-in-DFT calculations to determine the binding energy of CO on the Cu(111) surface.\cite{Welborn:QE-proj-COonMe:2018}

In the embedded EOM-EA-CCSD calculations, we partitioned both clusters into a high-level EOM-EA-CCSD fragment, including HCl and four of the gold atoms, shown in red in Figure \ref{fig:Au-clustermodels},
and a low-level DFT fragment, representing the remaining gold atoms of the surface cluster model, shown in yellow in Figure \ref{fig:Au-clustermodels}.
For the smaller HCl/Au\(_{10}\) cluster, it was possible to compare the EOM-EA-CCSD-in-DFT calculations with all-atom EOM-EA-CCSD results, whereas, for the larger HCl/\ch{Au38} cluster, calculations were only possible using EOM-EA-CCSD-in-DFT. 

For all EOM-EA-CCSD-in-DFT calculations, we employed the long-range corrected CAM-B3LYP\cite{Yanai:CAMB3LYP} functional as the low-level method, which is expected to provide an improved description of the electron-attached states.\cite{Vydrov:LR-DFT:2007,Jensen:LR-DFT:2010,Parravicini:QE-proj-EOM-CC:2021} 
In addition, for comparison with the periodic PBE results for HCl on Au(111), we performed PBE single--point calculations on the HCl/\ch{Au10} and HCl/\ch{Au38} clusters. 
When computing PBE adsorption energies for HCl/\ch{Au10} and HCl/\ch{Au38} using Gaussian-type orbitals as basis set, these energy differences were corrected for the basis set superposition error.
 
Besides energies, we computed Dyson orbitals of the HCl/\ch{Au38} anionic states at the EOM-EA-CCSD-in-CAM-B3LYP level. 
Dyson orbitals can be thought of as transition density matrices between the neutral and electron-attached states, providing a visual representation of the correlated orbital in which the additional electron resides.\cite{Melania:Dyson:2007,Jagau:Dyson:2016,Krylov:Orbitals:2020}

In addition, to compare the dissociation of HCl in the gas phase with our results for HCl on Au(111), we computed the dissociation curve of HCl using the spin-flip variant\cite{Krylov:SF:2001,Krylov:SF:2006,Casanova:SF:2020} of EOM-CCSD (EOM-SF-CCSD). 
For the EOM-SF-CCSD calculations, triplet reference states were computed using restricted open-shell Hartree-Fock. 
Spin-flip approaches are well-suited for describing bond breaking in gas phase molecules.\cite{Krylov:SF:2001,Krylov:SF:2006,Casanova:SF:2020}
On the other hand, for the cluster models, embedded CCSD is a valid approach for treating the dissociative adsorption of neutral HCl on Au(111) as the wavefunction does not exhibit multiconfigurational character.  
For the HCl/\ch{Au38} cluster, the norm of the $T_2$ amplitudes remains nearly constant along the reaction coordinate, increasing from 0.51 at the equilibrium structure to 0.53 at dissociation.
%is relatively small, 0.51 at the equilibrium structure, and, more importantly, this value does not significantly change along the reaction coordinate, being 0.53 at dissociation. 

In all cluster calculations, for hydrogen and chlorine, the basis sets used were aug-cc-pVTZ.\cite{H_basis_ccpVTZ,H_aug_ccpVTZ,Cl_aug_ccpVTZ}
For gold atoms, the LANL2TZ effective core potential (ECP) and basis set were used\cite{gold_ecp_lanl,gold_basis_lanl}.
Within this basis, the [Kr] \(4\mathrm{d}^{10} 4\mathrm{f}^{14}\) electrons are treated with the ECP and the \(5\mathrm{s}^2 5\mathrm{p}^6 5\mathrm{d}^{10} 6\mathrm{s}^1\) electrons are considered explicitly. 
In all correlated calculations, core electrons were kept frozen, correlating only the 1$\mathrm{s}^1$ electron for H, the 3$\mathrm{s}^2$3$\mathrm{p}^5$ electrons for Cl, and the 5$\mathrm{d}^{10}6\mathrm{s}^{1}$ electrons for Au. 
For the gas-phase HCl molecule, we also investigated the effect of using the aug-cc-pVQZ and aug-cc-pV5Z basis sets on the EOM-SF-CCSD dissociation energies. 
The dissociation energy increased by 0.06 eV when the basis set was changed from aug-cc-pVTZ to aug-cc-pVQZ, and by an additional 0.02 eV when transitioning from aug-cc-pVQZ to aug-cc-pV5Z. 
The EOM-SF-CCSD dissociation curves and energies are presented in Figure S5 and Table S1 of the SI.
Furthermore, within our embedding scheme, to partition the occupied orbitals between the low-level and high-level fragments, orbitals were first localized using the Pipek-Mezey method\cite{Pipek:loc:1989} and then assigned to each fragment based on Mulliken population analysis.\cite{Mulliken:1955}
Moreover, we truncated the virtual orbital space using concentric localization.\cite{Claudino:TruncationVirtual:2019} 
All calculations for the HCl/Au\(_{10}\)  and HCl/Au\(_{38}\) cluster models were conducted using a locally modified copy of the {\sl Q-Chem} software, version 6.0.2, using the implementation of ECPs within projection-based quantum embedding included in the 6.1.0 release of the program.\cite{QChem5}

\section{Results and discussion}
\protect\label{sec:Results}

\subsection{Periodic and cluster DFT results}

\begin{figure*}[h!]
    \includegraphics[width=16cm]{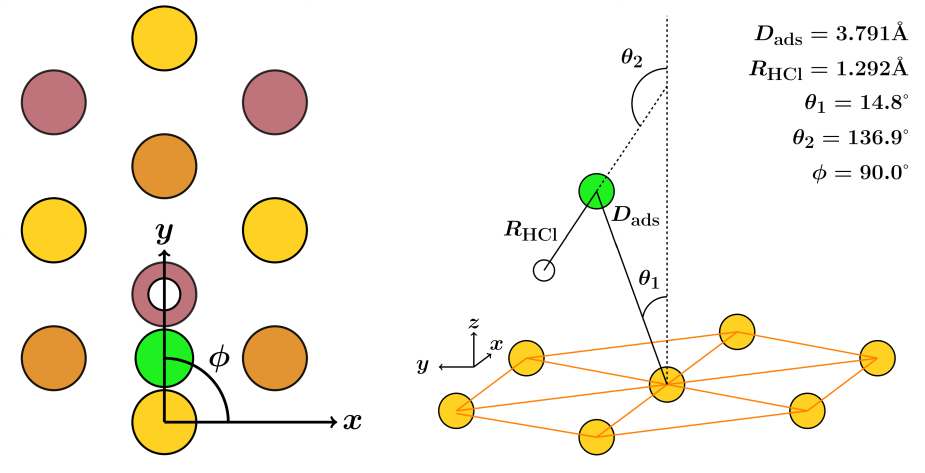}
    \centering
    \caption{Top (left) and side (right) views of the adsorption site of HCl on Au(111). Color code: H -- white, Cl -- green, Au layer 1 -- yellow, Au layer 2 -- orange, Au layer 3 -- purple. On the Au(111) surface, Cl occupies the atop site and H the hollow-\emph{fcc} site.}
    \protect\label{fig:ads-site}
\end{figure*}
For the bulk gold, the optimized PBE lattice constant is 4.154 {\AA}, resulting in a bulk Au--Au distance of 2.937 {\AA}. 
For the Au(111) surface, the inter-layer distances are found to be $d_{12}$ = 2.411 {\AA}, $d_{23}$ = 2.377 {\AA}, and $d_{34}$ = 2.411 {\AA}, with layers numbered from top to bottom. 
Upon HCl adsorption on Au(111), these inter-layer distances exhibit minimal changes, with differences within 2 pm, indicating only minor surface relaxation. 
Figure \ref{fig:ads-site} illustrates the adsorption site of HCl on Au(111).
In this configuration, the hydrogen atom of HCl points toward the metal surface relative to the chlorine atom, with the hydrogen and chlorine atoms occupying the hollow-\emph{fcc} and atop adsorption sites, respectively. 
Furthermore, following the adsorption of HCl onto Au(111), the stretching frequency of HCl undergoes a red shift from 2899 to 2808 cm\(^{-1}\), corresponding to a lengthening of the H--Cl bond from 1.286 to 1.292 {\AA}. 
This indicates a certain degree of charge transfer from the metal surface to the $\upsigma^{\ast}$  antibonding orbital of HCl, leading to the weakening of the H--Cl bond.

\begin{table}[h!]
    \centering
    \caption{PBE+D3 adsorption energies $\Delta E_{\text{ads}}$ in eV for HCl/Au(111) periodic and cluster models and long-range corrections $\Delta_{\text{LR}}$ in eV. All calculations are performed at the periodic PBE equilibrium structure of HCl/Au(111). Adsorption energies for the clusters are corrected for the basis set superposition error.} 
    \protect\label{tab:adsorption-energies}
    \begin{tabular*}{\textwidth}{l@{\extracolsep\fill}cccr}
     \hline 

     & \multicolumn{3}{c}{$\Delta E_{\text{ads}}$} & $\Delta_{\text{LR}}$\\
             \cline{2-4} 
     System        & PBE & D3 & PBE+D3 & \\
     \hline 
     HCl/Au (pbc)& $-$0.056 &  $-$0.201 & $-$0.257 & 0.000 \\
     HCl/\ch{Au10}& $-$0.047 &  $-$0.121 & $-$0.167 & $-$0.089 \\
     HCl/\ch{Au38} & $-$0.049 &  $-$0.152 & $-$0.201 & $-$0.056 \\
    \hline
    \end{tabular*}
\end{table}
Table \ref{tab:adsorption-energies} presents the PBE+D3 adsorption energies for periodic HCl/Au(111), and the HCl/\ch{Au10} and HCl/\ch{Au38} cluster models.
The adsorption energy, $\Delta E_{\text{ads}}$, is defined by the equation:
	\begin{equation}
	  \label{eq:ads}
	\Delta E_{\text{ads}} = E_{\text{HCl/Au}} - (E_{\text{Au}} + E_{\text{HCl}}),  
	\end{equation}
where $E_{\text{HCl/Au}}$ is the energy of the adsorbate-surface system, $E_{\text{Au}}$ is the energy of the slab, and $E_{\text{HCl}}$  corresponds to the energy of the HCl molecule in the gas phase. 
The dispersion term, computed using the D3 correction\cite{Grimme:D3:2010} to PBE, accounts for 80\% of the total adsorption energy, which means that HCl only weakly physisorbs to the Au(111) surface.

\begin{figure*}[!h]
    \centering
    \includegraphics[width=16cm]{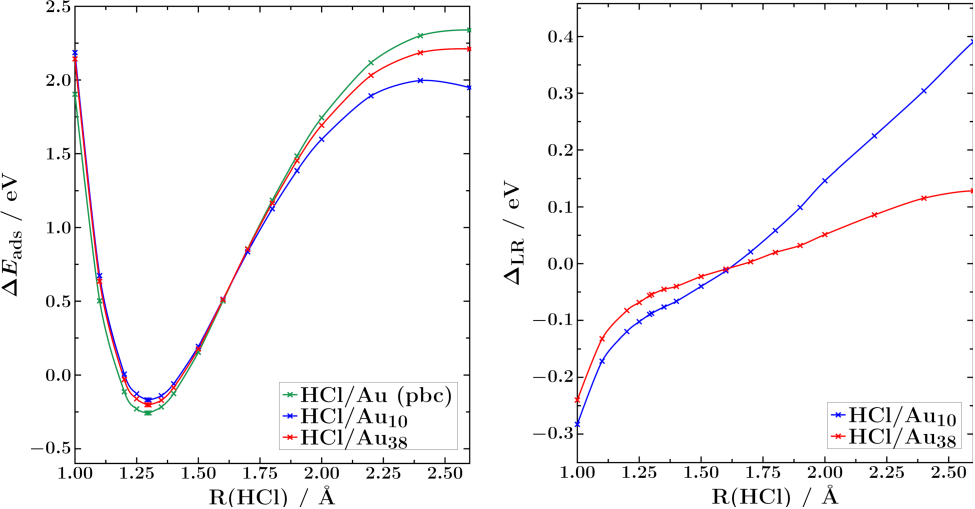}
    \caption{Left: PBE + D3 potential energy curves for HCl/Au(111) periodic  (green curve), HCl/\ch{Au10} (blue curve), and HCl/\ch{Au38} (red curve). Right: Long-range correction $\Delta_{\mathrm{LR}}$ for HCl/Au\(_{10}\) and HCl/Au\(_{38}\) clusters as a function of the H--Cl bond length. Adsorption energies for the clusters are corrected for the basis set superposition error.}
    \label{fig:long-range-correction}
\end{figure*}

Using finite-size models for metals can introduce artifacts and lead to slow convergence of the calculated energies and properties with the cluster size.\cite{Huang:ClusterLimitations:2008}
In light of this, we examined the convergence of the PBE+D3 adsorption energy across two clusters with varying size and geometry. 
At the PBE equilibrium adsorption structure, as the cluster size increases from HCl/\ch{Au10} to HCl/\ch{Au38}, the long-range correction,\cite{Alessio:CH4:2018,Alessio:H2O:2019} 
\begin{equation}
	\Delta_{\text{LR}} = \Delta E_{\text{ads}} (\text{pbc}) -  \Delta E_{\text{ads}} (\text{cluster}),  
\end{equation}
which is defined as the difference in adsorption energy between the periodic (pbc) and cluster models, decreases by 0.03 eV in absolute value.
The HCl/\ch{Au38} cluster recovers 80\% of the adsorption energy computed with the periodic model, while the smaller HCl/\ch{Au10} cluster captures only about 60\%, indicating that the larger cluster better accounts for the long-range contribution to the binding.
Figure \ref{fig:long-range-correction} shows the PBE+D3 dissociation curves for the HCl on Au(111) periodic and cluster models, along with the long-range corrections for HCl/\ch{Au10} and HCl/\ch{Au38} as a function of the H--Cl bond length. 
Interestingly, as the H--Cl bond length extends beyond 1.75 Å, we observe a transition in the long-range correction from negative to positive values.
This transition is accompanied by a steeper increase in $\Delta_{\text{LR}}$ for HCl/\ch{Au10} as the HCl dissociation progresses.
This occurs because the \ch{Au10} cluster is too small to bind the hydrogen atom as the H--Cl bond stretches beyond 2 {\AA}, resulting in larger deviations of the PBE dissociation curve for HCl/\ch{Au10} (blue curve) compared to HCl/\ch{Au38} (red curve) from the periodic one (green curve). 

Furthermore, we note that even at the PBE level, and regardless of the cluster size and geometry, employing a finite model for a metallic surface results in non-zero Mulliken charges on specific metal atoms and introduces an artificial HOMO-LUMO gap. 
For instance, in the case of \ch{Au38}, we observe a HOMO-LUMO gap of 0.3 eV. 
Additionally, metal atoms located at the center of \ch{Au38} exhibit non-zero Mulliken charges, yet the surface cluster model maintains its charge neutrality.

\begin{figure}[!h]
    \centering
    \includegraphics[width=8cm]{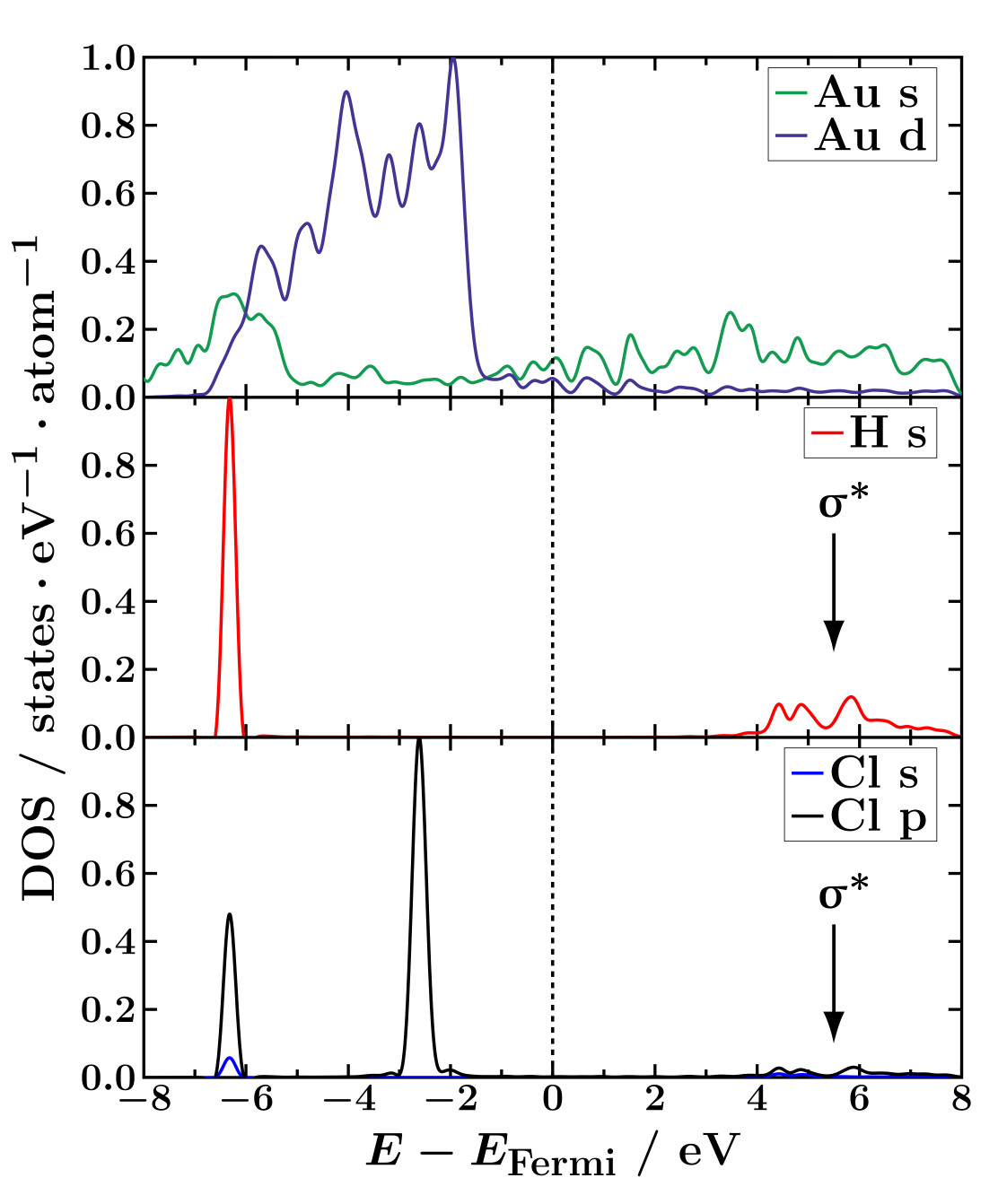}
    \caption{Atomic orbital projected density of states (DOS) of HCl/Au(111) at the equilibrium structure, computed using PBE and normalized such that the maximum values are set to 1.}
    \protect\label{fig:DOS_HClonGold}
\end{figure}
Figure \ref{fig:DOS_HClonGold} illustrates the electronic density of states (DOS) projected onto the Au 5d and 6s, H 1s, and Cl 3s and 3p atomic orbitals, computed at the HCl/Au(111) equilibrium structure.
When HCl adsorbs onto the gold surface, its $\upsigma^{\ast}$ antibonding orbital spreads over a wide energy range, between 4 and 6 eV above the Fermi level, mainly due to hybridization with the Au s-band electrons. 
This DOS analysis indicates that attaching an additional electron to the antibonding orbital of HCl on Au(111) requires a significant amount of energy, suggesting that gold nanoparticles would require a plasmon resonance energy greater than 4 eV to mediate the cleavage of the HCl bond.
At the HCl/Au(111) equilibrium structure, we also computed the Bader charges for HCl, the gold surface, and HCl/Au(111).
The corresponding values are listed in Table S2. 
For the HCl molecule in the gas phase, the Bader charges are $+$0.348 a.u. for the hydrogen atom and $-$0.348 a.u. for the chlorine atom. 
Upon adsorption, the Bader charges shift to $+$0.337 a.u. for H and $-$0.353 a.u. for Cl. 
This reflects a total charge transfer of $-$0.016 a.u. from Au(111) to HCl.
This small charge transfer is consistent with the large HOMO-LUMO gap predicted by the DOS.
However, relying on the periodic DOS is analogous, in the molecular domain, to considering the relative energies of the HOMO and LUMO frontier molecular orbitals, which often only provides a limited and sometimes qualitatively incorrect picture of electronic excitation, ionization, and electron attachment.
In the following section, by employing embedded EOM-EA-CCSD, we go a step further, and directly compute electron attachment energies as energy differences between the charged and neutral states of HCl on Au(111). 

\subsection{Embedded EOM-CCSD results}

At the PBE equilibrium adsorption structure, we employed the smaller HCl/\ch{Au10} cluster to assess the performance of embedded EOM-EA-CCSD with respect to all-atom EOM-EA-CCSD.
However, the dissociation curves are not reported because this smaller cluster cannot accommodate the HCl molecule at elongated H--Cl bond distances.
We used the larger HCl/\ch{Au38} cluster to compute embedded EOM-EA-CCSD dissociation curves for both neutral and electron-attached states of HCl on Au(111).

\begin{figure}
    \centering
    \includegraphics[width=8cm]{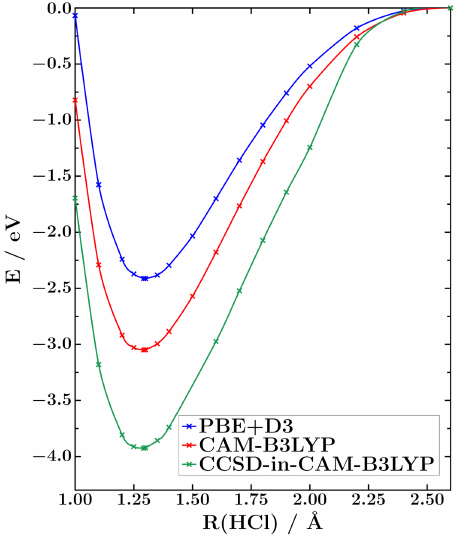}
    \caption{PBE+D3, CAM-B3LYP and CCSD-in-CAM-B3LYP potential energy curves of HCl/\ch{Au38}. All energies relative to the dissociation limit at that level of theory.}
    \label{fig:embed-pbe-cam}
\end{figure}

Figure \ref{fig:embed-pbe-cam} presents a comparison of the dissociation curves of the neutral HCl on \ch{Au38} obtained using PBE+D3, CAM-B3LYP, and CCSD-in-CAM-B3LYP.
The corresponding dissociation energies are listed in Table \ref{tab:everything}.
Although CAM-B3LYP represents an improvement over PBE+D3, both DFT calculations underestimate the dissociation energy compared to embedded CCSD, with corrections of 0.88 eV for CAM-B3LYP and 1.52 eV for PBE+D3. 
To compare the dissociation of neutral HCl on Au(111) to that of HCl in the gas phase, we use CCSD-in-CAM-B3LYP dissociation curves for HCl/Au(111) and an EOM-SF-CCSD curve for the HCl molecule (Figure S5).  
From this comparison, the dissociation energy of the HCl molecule reduces by 0.55 eV upon adsorption on Au(111).  
\begin{table}[h!]
	\centering
	\caption{Dissociation energies $D_e$ in eV and electron-attachment energies $\Delta E_{\mathrm{EA}}$ in eV for HCl, as well as for HCl upon adsorption on Au(111).} 
        \protect\label{tab:everything}
	\begin{tabular*}{\textwidth}{lll@{\extracolsep\fill}c}
	\hline
	System & &Method & $D_e$ \\
	\hline
	\\
	[-0.6em]
	HCl & &EOM-SF-CCSD & 4.48$^b$ \\
	\\
	[-0.6em]
	\multirow{3}{*}{HCl/\ch{Au38}} & & PBE+D3 & 2.41\\
	{} & {} & CAM-B3LYP & 3.05\\
	{} & {} & CCSD-in-CAM-B3LYP & 3.93\\
	\\
	[-0.6em]
	\multirow{2}{*}{HCl$^{-}$/\ch{Au38}} &State 1 & EOM-EA-CCSD-in-CAM-B3LYP & 3.44\\
	{}&State 5 & EOM-EA-CCSD-in-CAM-B3LYP & 3.21\\
	\\
	 & & & $\Delta E_{\mathrm{EA}}$\\
	\hline
	\\[-0.6em]
	\multirow{2}{*}{HCl/\ch{Au10}} & \multirow{2}{*}{State 3} &EOM-EA-CCSD & $-1.73$\\
	{} & & EOM-EA-CCSD-in-CAM-B3LYP & $-1.47$\\
	\\[-0.6em]
	\multirow{2}{*}{HCl/\ch{Au38}} & State 1 & EOM-EA-CCSD-in-CAM-B3LYP & $-2.25$\\
	{} & State 5 & EOM-EA-CCSD-in-CAM-B3LYP & $-0.89$\\
	\hline
	\end{tabular*}
	
	$^a$ All $\Delta E_{\mathrm{EA}}$ values are computed at the PBE equilibrium adsorption structure.
	$^b$ $D_e$ is derived from the EOM-SF-CCSD dissociation curve of HCl, which is shown in Figure S5. 
\end{table}

\begin{figure*}[h!]
    \centering
    \includegraphics[width=16cm]{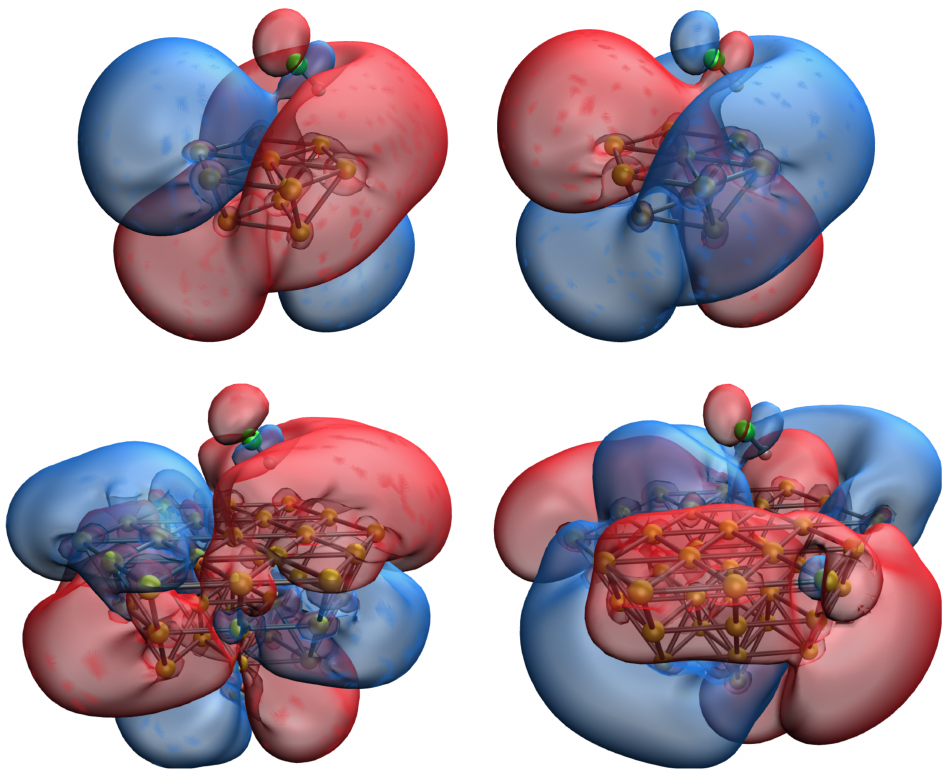}
    \caption{Top: Dyson orbitals of state 3 for HCl$^-$/\ch{Au10} computed using EOM-EA-CCSD (left)  and EOM-EA-CCSD-in-CAM-B3LYP (right). Bottom: Dyson orbitals of states 1 (left) and 5 (right) for HCl$^-$/\ch{Au38} computed using EOM-EA-CCSD-in-CAM-B3LYP. An isovalue of 0.002 was used.}
    \label{fig:cluster-dysons}
\end{figure*}

At the PBE equilibrium adsorption structure, we compute a manifold of six bound anionic states for HCl/\ch{Au10} and nine for HCl/\ch{Au38} using embedded EOM-EA-CCSD. 
%Their corresponding electron-attachment energies are shown in the left panel of Figure \ref{fig:eom-cc-in-camb3lyp}.
Only some of these states correspond to Kohn-Sham virtual orbitals with the extra electron displaying $\upsigma^{\ast}$ antibonding character on HCl.
We identify one dissociative electron-attached state for HCl/\ch{Au10}, the third lowest in energy, and two for HCl/\ch{Au38}, the first and fifth lowest in energy.
The associated Kohn-Sham virtual orbitals are illustrated in Figures S6 and S7, while
Table \ref{tab:everything} reports their electron-attachment energies.
For HCl/\ch{Au10}, both embedded EOM-EA-CCSD and full EOM-EA-CCSD yield only one electron-attached state of dissociative character,  though the corresponding electron-attachment energy differs by 0.26 eV between the two methods.
For HCl/\ch{Au38}, the electron-attached states 1 and 5 span an energy range of 1.36 eV, with the energy of the electron-attached state 3 of HCl/\ch{Au10} falling roughly in the middle.
%For these states electron-attached states, EOM-CCSD single-excitation $R_1$ amplitudes are 0.99, and 99\% of this single-electron attachement is to a single $\upsigma^*$-like molecular orbital.
For these states, electron attachment to one single $\upsigma^{\ast}$-like orbital accounts for 98\% of the EOM-CCSD eigenvector.
Figure \ref{fig:cluster-dysons} shows the Dyson orbitals for attachment to these states.
Based on this orbital picture, the extra electron attaches to hybridized orbitals of HCl/Au(111), arising from the interaction between the $\upsigma^{\ast}$ antibonding orbital of HCl and the s and d conduction bands of the metal surface. 
These orbitals feature a nodal plane between the hydrogen and chlorine atoms, characteristic of the HCl $\upsigma^{\ast}$ antibonding orbital, and show a bonding interaction between the hydrogen and the gold surface. 
Additionally, they have diffuse character on the surface atoms, indicating the tendency of the extra charge to delocalize to the metal.
By projecting the Dyson orbitals of the electron-attached states of HCl/\ch{Au38} onto the atoms, we observe that HCl contributes solely 0.1\% to the orbital of both states 1 and 5, while the four gold atoms of the EOM-EA-CCSD fragment contribute 38.2\% for state 1 and 63.8\% for state 5.
For state 3 of HCl/\ch{Au10}, Dyson orbitals calculated using embedded EOM-EA-CCSD and all-atom EOM-EA-CCSD closely match, as illustrated in Figure \ref{fig:cluster-dysons}.
Furthermore, the increase in the number of anionic states from six to nine, and the number of dissociative electron-attached states from one to two, as the cluster size increases from HCl/\ch{Au10} to HCl/\ch{Au38}, better describes the $\upsigma^{\ast}$ antibonding band of anionic states expected in the periodic HCl$^-$/Au(111) system. 
\begin{figure*}[h!]
    \centering
    \includegraphics[width=16cm]{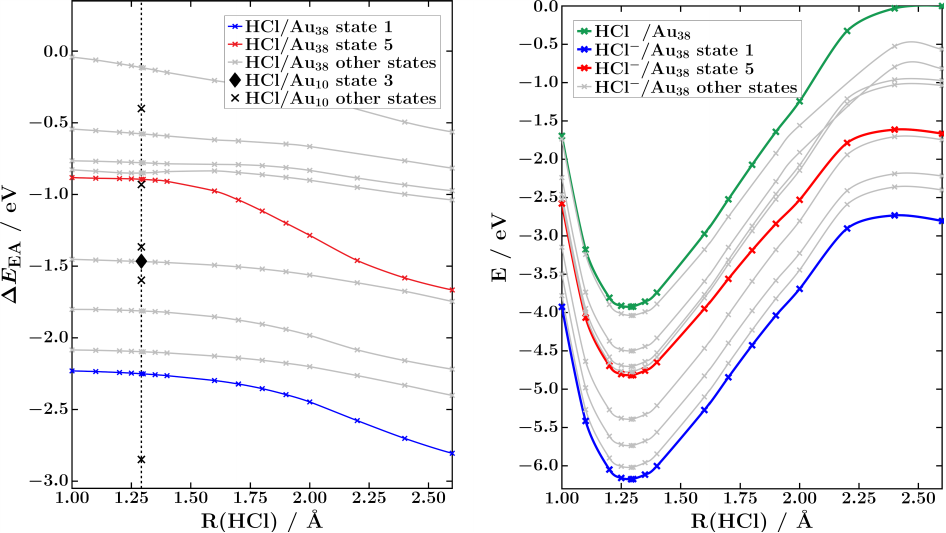}
    \caption{Left: Attachment energies $\Delta E_{\text{EA}}$ for HCl/\ch{Au38} computed using EOM-EA-CCSD-in-CAM-B3LYP as a function of the H--Cl bond length. 
    $\Delta E_{\text{EA}}$ values for the anionic states 1 and 5 of HCl/\ch{Au38} are in blue and red, respectively, with the remaining values in gray. 
    Black crosses indicate $\Delta E_{\text{EA}}$ for HCl/\ch{Au10} computed using EOM-EA-CCSD-in-CAM-B3LYP at the PBE equilibrium adsorption structure, with the diamond symbol highlighting state 3. 
    Right: EOM-EA-CCSD-in-CAM-B3LYP potential energy curves of the neutral ground state (green) and anionic states 1 (blue) and 5 (red) of HCl/\ch{Au38}.}
    \label{fig:eom-cc-in-camb3lyp}
\end{figure*}

Furthermore, for HCl/\ch{Au38}, we investigated the variation in the attachment energy for all nine bound anionic states as a function of the H--Cl bond length, i.e. $\Delta E_{\mathrm{EA}}$ vs. $R$(HCl). 
The left panel of Figure \ref{fig:eom-cc-in-camb3lyp} presents the results, while the right panel shows the corresponding EOM-EA-CCSD-in-CAM-B3LYP dissociation curves.
The deviation of the $\Delta E_{\mathrm{EA}}$ vs. $R$(HCl) curve from a horizontal line indicates a reduction in the dissociation energy of HCl$^-$ on \ch{Au38} relative to HCl on \ch{Au38}, implying antibonding character of the corresponding anionic state. 
When the electron is attached to orbitals of $\upsigma^*$ character, rather than other virtual orbitals, the dissociation energy lowers. 
This means that the dissociation curves of HCl and \ch{HCl-} on \ch{Au38} do not run parallel and thus, the electron attachment energy is more negative at stretched geometries. 
These curves for the anionic states 1 and 5 of HCl$^-$/\ch{Au38}, shown in blue and red, respectively, clearly deviate from linearity as the H--Cl bond stretches.
From the corresponding dissociation curves, we computed dissociation energies for states 1 and 5 that reduce by 0.49 and 0.72 eV, respectively, compared to that for the neutral state. 
This is on average a reduction of 0.61 eV.
This result is corroborated by the Dyson orbitals for the anionic states 1 and 5, which illustrate partial occupancy of the HCl $\upsigma^{\ast}$ antibonding orbital.
Additionally, the $\Delta E_{\mathrm{EA}}$ vs. $R$(HCl) curves for states 3 and 9 also show a deviation from linearity; however, the corresponding electron attachment does not involve orbitals with HCl $\upsigma^{\ast}$ antibonding character (Figure S8). 
Thus, reaction pathways involving electron-attached states 3 and 9 are expected to contribute less to the dissociation of HCl$^{-}$ on Au(111).
Notably, all nine anionic states are bound at the equilibrium adsorption structure and remain so as the H--Cl bond breaks, making attachment of an extra charge easier and permanent. 
This contrasts with gas-phase HCl, which does not support any bound anionic states at its equilibrium distance, only becoming bound when the bond elongates.\cite{Nesbet:VirtualState:1977,Cizek:VirtualHCl:1999,Hotop:VirtualState:2003,Fedor:VirtualHCl:2010,Moorby:HCl:2024}
Furthermore, the diffuse nature of the Dyson orbitals on the metal surface suggests that the delocalization of the excess electron to the metal drives the stabilization of the anion compared to the gas phase.
%We also observe that the computed dissociation curves of HCl on \ch{Au38} do not exhibit any cusp or discontinuity, unlike other cases where dramatic geometrical changes during a chemical reaction complicate the localized orbital partitioning of the system into fragments along the reaction coordinate.\cite{Welborn:QE-proj-COonMe:2018}
%We attribute this smooth behavior to the fact that the the reactant HCl remains within the EOM-CCSD fragment across all bond lengths.

\section{Conclusions}

We investigated the dissociative electron attachment of HCl adsorbed on Au(111) using projection-based EOM-EA-CCSD-in-DFT embedding applied to cluster models of the periodic HCl/Au(111) system. 
Our findings reveal that populating the $\upsigma^{\ast}$ antibonding band of HCl on Au(111) can significantly enhance HCl dissociation compared to neutral HCl in the gas phase, lowering the dissociation energy by 1.16 eV. 
From the Dyson orbital of the electron-attachment process, we attribute the stabilization of HCl$^-$ upon adsorption on Au(111) to the delocalization of the additional charge to the metal surface.
This results in the anion being bound at all H--Cl bond lengths.
These findings underscore the feasibility of plasmon-induced dissociation of HCl on gold nanoparticles through bound anions. 
Furthermore, from the PBE density of states, the band of HCl antibonding orbitals lies well above the Fermi level, whereas embedded EOM-CCSD yields bound anionic states, 
underscoring the importance of using correlated methods to accurately model charged species on surfaces.

The HCl anion in the gas phase is unbound near the HCl equilibrium bond length.\cite{Nesbet:VirtualState:1977,Cizek:VirtualHCl:1999,Hotop:VirtualState:2003,Fedor:VirtualHCl:2010,Moorby:HCl:2024} 
Moreover, the low energy electron scattering from neutral HCl is characterized by significantly enhanced cross-sections at threshold energies.\cite{Schafer:HClExpt:1991,Hotop:VirtualState:2003} 
Under the stabilizing effect of the metal surface, HCl$^-$ can easily become bound. 
In contrast, other diatomic molecules like H$_2$ and N$_2$ have valence unbound electron-attached states with energies well above those of their respective neutral molecules.\cite{Jagau:N2H2:2014,White:N2H2:2017} 
Plasmon-driven dissociation of H$_2$ and N$_2$ on gold nanoparticles has been already investigated.\cite{MukherjeeCarter:PlasmonExptTheory:2012,Martinez:N2:2016}
While the presence of a metal surface will stabilize the anionic states of H$_2$ and N$_2$, it may not be enough to turn them into stable anions, as in HCl.
Further applications of embedded EOM-EA-CCSD to H$_2$ and N$_2$ on gold systems are needed, where unbound anions, rather than bound ones, may play a role in the catalysis. 

Furthermore, we observed that using a cluster approach to model metal surfaces can introduce deficiencies already present at the DFT level, such as non-zero Mulliken charges and non-zero HOMO-LUMO gaps. 
The impact of these shortcomings on the computed electron-attachment and dissociation energies is difficult to quantify.
In spite of this, cluster-based wavefunction-in-DFT embedding schemes have been shown to describe binding of small molecules on metal surfaces accurately,\cite{Welborn:QE-proj-COonMe:2018} and projection-based embedding schemes have been used to describe various types of excited states.\cite{Bennie:QM-proj-exc:2017,Parravicini:QE-proj-EOM-CC:2021,Alessio:QE-proj-OS:2024} 
In this work we have combined for the first time the strengths of both approaches and described electron-attached states at metal surfaces.
We anticipate that further advancements in our computational scheme, such as the incorporation of periodic embedding potentials built from localized Wannier orbitals,\cite{Libisch:Warnier:2017,Schaefer:LoclizedOrbitals:2021} will help to improve the representation of the metallic system. 
This refinement of the computational approach will be key for the continued application of embedded EOM-CCSD methods in surface science. 

\section*{Conflicts of interest}
There are no conflicts to declare.

\section*{Acknowledgement}
The authors thank Prof. Andreas Gr\"{u}neis (TU Wien) for helpful conversations on the physics of molecular adsorption and dissociation at metal surfaces.
R.E.M. acknowledges the European Union for mediating a bilateral exchange as a part of the Erasmus+ program.
This work has been supported by the Marie Sk\l{}odowska-Curie Actions fellowship to M.A. (Grant Agreement No. 101062717).
T.-C.J. gratefully acknowledges funding from the European Research Council (ERC) under the European Union’s Horizon 2020 research and innovation program (Grant Agreement No. 851766) and the KU Leuven internal funds (Grant No. C14/22/083).

\section*{Supplementary Information}
The following files are available free of charge.
\begin{itemize}
  \item SI-HClonGold.pdf: Convergence tests for DFT periodic calculations, EOM-SF-CCSD dissociation curve of HCl, Bader charges, additional Dyson and Kohn-Sham orbitals, relevant Cartesian coordinates.
\end{itemize}

\bibliography{HClonGold.bib}
	\bibliographystyle{rsc}
\newpage
%
%\section*{TOC Graphic}
%\begin{figure}[h!]
%\centering
%\includegraphics[width=8.25cm]{toc.pdf}
%\end{figure}

\end{document}

\title{
 Dissociative Electron Attachment on Metal Surfaces: \\
 The Case of \ch{HCl-} on Au(111)
 }
\author{R. E. Moorby,$^{\dagger,\ddagger}$ V. Parravicini,$^\dagger$ T.-C. Jagau,$^\dagger$ and M. Alessio$^{\ast,\dagger}$\\
{\small $^\dagger$Department of Chemistry, KU Leuven, Celestijnenlaan 200F, B-3001 Leuven, Belgium}\\
{\small $^\ddagger$Department of Chemistry, University of Durham, South Road, Durham, DH1 3LE, United Kingdom}\\
{\small $^{\ast}$ E-mail: maristella.alessio@kuleuven.be}}
\date{}

\maketitle

\tableofcontents

\clearpage

\section{Convergence tests for the DFT periodic calculations}
\protect\label{sec:Convergence}

\begin{figure}[h]
     \centering
     \includegraphics[width=7.1cm]{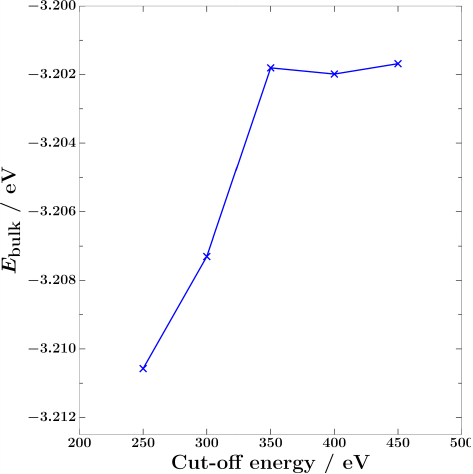}
     \caption{Energy of the gold fcc bulk ($E_{\mathrm{bulk}}$ in eV) with respect to the energy cut-off of the plane-wave basis set.}
     \label{fig:encut-au} 
\end{figure}

\begin{figure}[h]
        \centering
        \includegraphics[width=7.1cm]{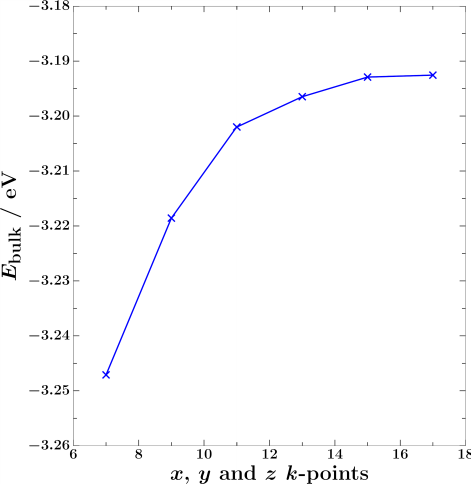}
        \caption{Energy of the gold fcc bulk ($E_{\mathrm{bulk}}$ in eV) with respect to the number of \emph{k}-points in each dimension of the \(\Gamma\)-centered $k$-point grid.}
        \label{fig:kpoints-au}
\end{figure}

\clearpage
    
\begin{figure}
        \centering
        \includegraphics[width=7.1cm]{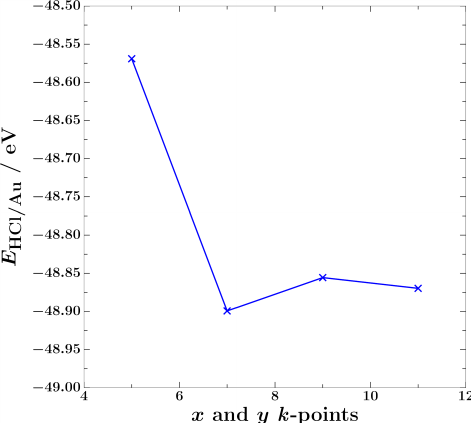}
        \caption{Energy of the HCl/Au(111) system ($E_{\mathrm{HCl/Au}}$ in eV) with respect to the number of \emph{k}-points in the $x$- and $y$-axes of the \(\Gamma\)-centered $k$-point grid.}
        \label{fig:kpoints-hclau}
\end{figure}
    
\begin{figure}
        \centering
        \includegraphics[width=7.1cm]{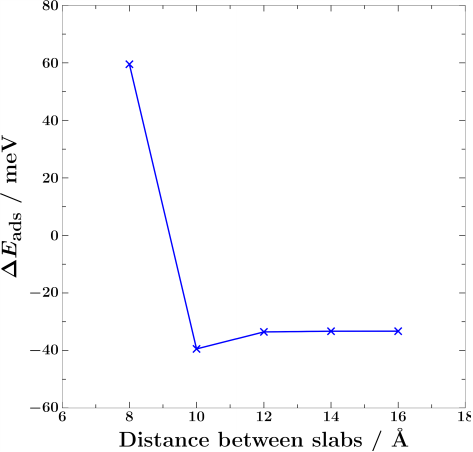}
        \caption{Adsorption energy ($\Delta E_\mathrm{ads}$) of HCl on Au(111) with respect to the distance between the periodically repeated slabs (along $z$-axis).}
        \label{fig:vacuum-hcl-au}
\end{figure}

\clearpage

\section{EOM-SF-CCSD dissociation curve of HCl}
\protect\label{sec:eom-sf}

\begin{figure}[h!]
    \centering
    \includegraphics[width=8cm]{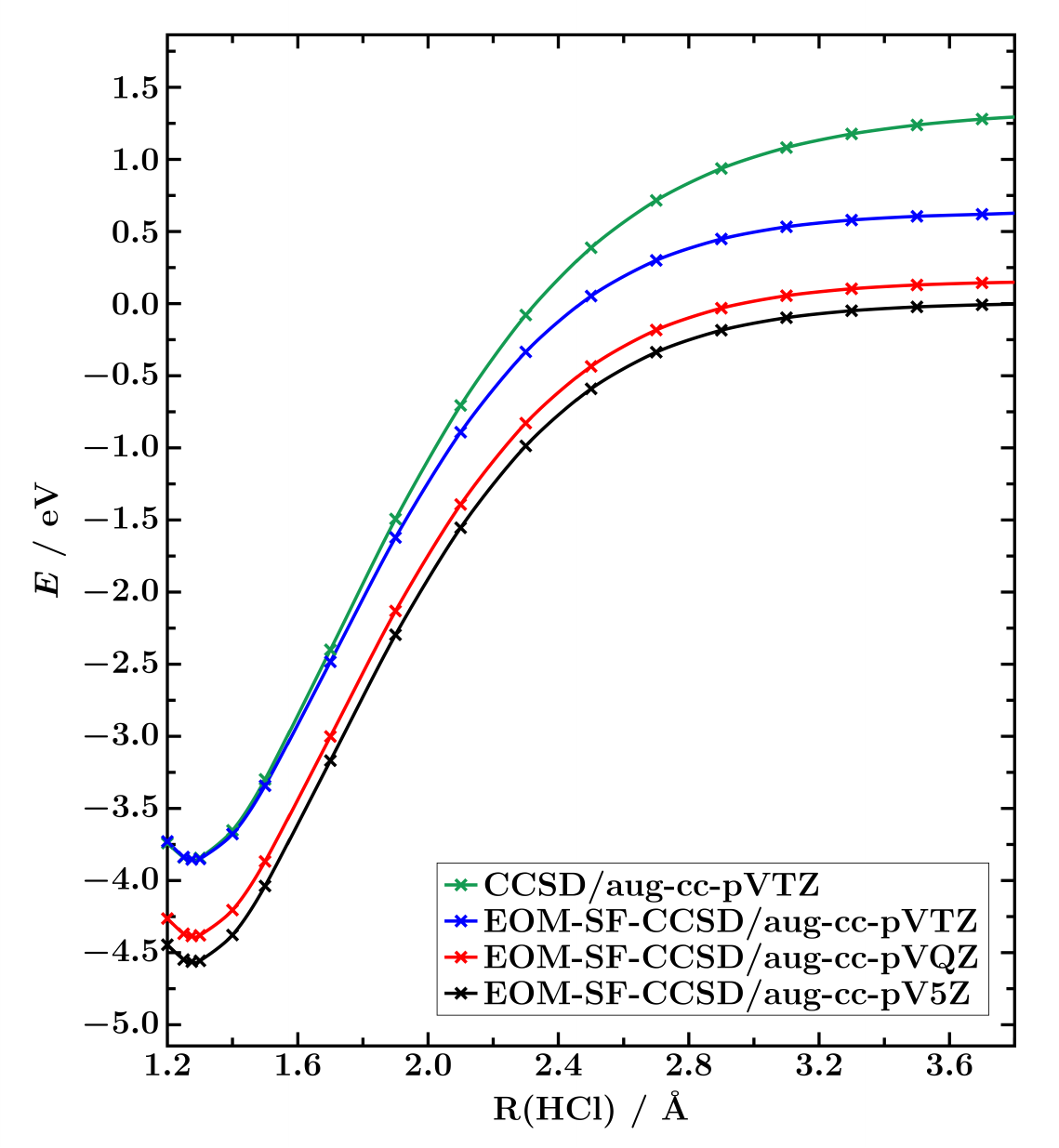}
    \caption{CCSD and EOM-SF-CCSD potential energy curves of HCl. CCSD curve is computed using aug-cc-pVTZ basis set. EOM-SF-CCSD curves are computed using aug-cc-pVXZ basis sets with X = T, Q, 5. All energies relative to the dissociation limit using EOM-SF-CCSD/aug-cc-pV5Z.}
    \label{fig:high-level-correction}
\end{figure}

\begin{table}[h!]
\caption{CCSD and EOM-SF-CCSD dissociation energies ($D_e$) of HCl in eV. CCSD dissociation energy is computed using aug-cc-pVTZ basis set. EOM-SF-CCSD dissociation energies are computed using aug-cc-pVXZ basis sets with X = T, Q, 5.}
\begin{tabular*}{\textwidth}{l@{\extracolsep\fill}lc}
\hline
Method & Basis set & $D_e$ \\
\hline
CCSD & aug-cc-pVTZ & 5.16 \\
EOM-SF-CCSD & aug-cc-pVTZ & 4.48 \\
EOM-SF-CCSD & aug-cc-pVQZ & 4.54 \\
EOM-SF-CCSD & aug-cc-pV5Z & 4.56 \\
\hline
\end{tabular*}
\label{tab:eom-sf}
\end{table}

\clearpage

\section{Bader charges}

\begin{table}[h!]
\caption{Bader charges (in atomic units) for HCl and the first-layer four Au atoms of Au(111), computed at the equilibrium structure of HCl/Au(111).}
\begin{tabular*}{\textwidth}{l@{\extracolsep\fill}cccc}
\hline
Atom & HCl$_{\mathrm{(g)}}$ & Au(111) & HCl / Au(111) & Effect of HCl \\
{} & {} & {} & {} & adsorption on Au(111) \\ 
\hline
 Au${(1)}$     &       & $-$0.030 & $-$0.041 & $-$0.011 \\
Au${(2)}$     &       & $-$0.031 & $-$0.042 & $-$0.011 \\
Au${(3)}$     &       & $-$0.034 & $-$0.025 & $+$0.009 \\
Au${(4)}$     &       & $-$0.024 & $-$0.005 & $+$0.019 \\
Au surface$^\mathrm{a}$    &       &       &       & $+$0.007 \\
H             & $+$0.348 &       & $+$0.337 & $-$0.011 \\
Cl            & $-$0.348 &       & $-$0.353 & $-$0.005 \\
HCl$^\mathrm{b}$             &       &       &       & $-$0.016 \\
\hline
\end{tabular*}
\\
$^\mathrm{a}$Sum of Bader charges in the first layer of Au(111). $^\mathrm{b}$Sum of Bader charges in HCl.
\label{tab:bader}
\end{table}

\clearpage

\section{Additional Dyson and Kohn-Sham orbitals}
\protect\label{sec:dyson-ks}

\begin{figure}[h!]
    \centering
    \includegraphics{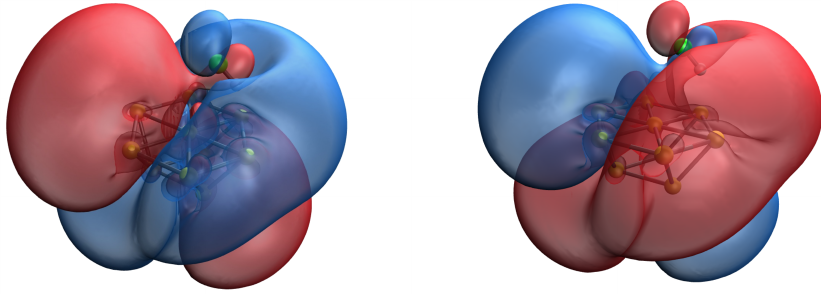}
    \caption{SCF orbitals involved 
    %(with leading amplitude, 95\% for EOM-EA-CCSD and 94\% for EOM-EA-CCSD-in-CAM-B3LYP) 
    in the electron-attachment transition to state 3 of HCl/Au$_{10}$. Left: HF orbital in all-atom EOM-EA-CCSD. Right: CAM-B3LYP orbital in EOM-EA-CCSD-in-CAM-B3LYP. An isovalue of 0.005 was used.} 
    \label{fig:SCF-Au10}
\end{figure}

\begin{figure}[h!]
    \centering
    \includegraphics{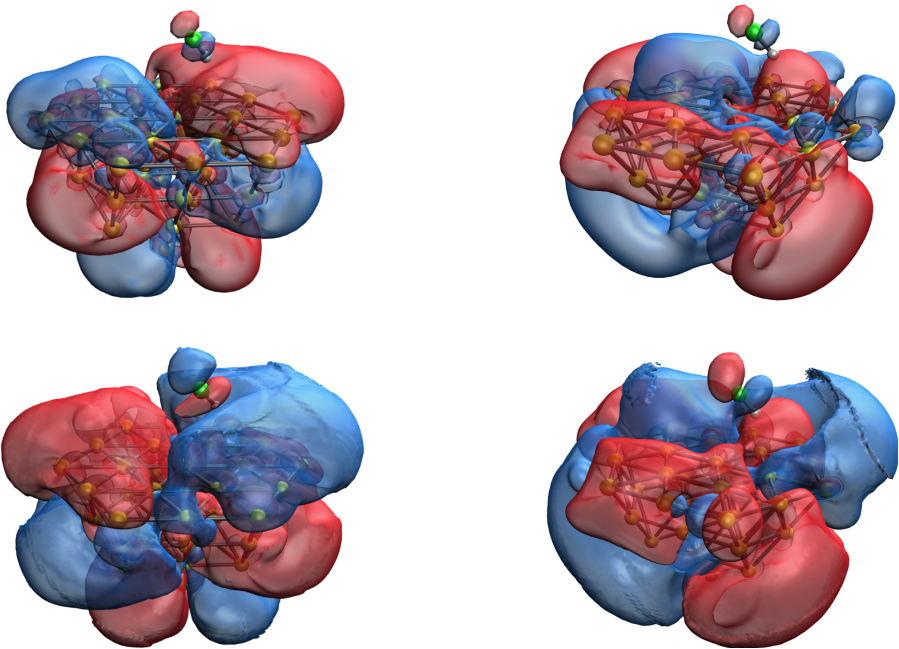}
    \caption{Top: Dyson orbitals of the electron-attached states 1 (left) and 5 (right) for HCl/Au$_{38}$ computed using EOM-EA-CCSD-in-CAM-B3LYP. Bottom: CAM-B3LYP Kohn-Sham orbitals involved 
    %(with leading amplitude, 99-98\%) 
    in the electron-attachment transitions to states 1 (left) and 5 (right) of HCl/Au$_{38}$. An isovalue of 0.005 was used.}
    \label{fig:dyson-ks}
\end{figure}

\begin{figure}[h!]
    \centering
    \includegraphics{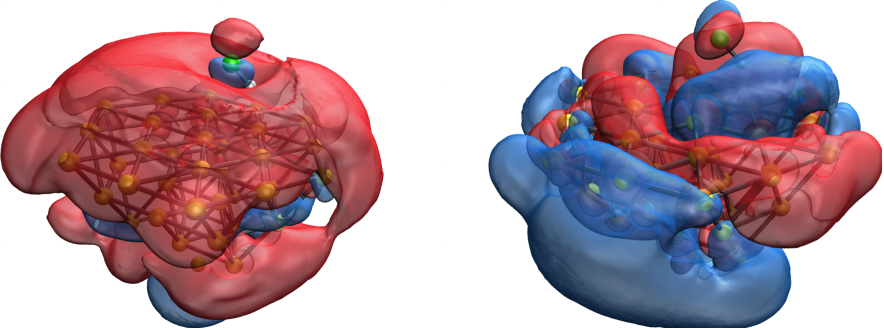}
    \caption{CAM-B3LYP Kohn-Sham orbitals involved %(with leading amplitude, 99-98\%) 
    in the electron-attachment transitions to states 3 (left) and 9 (right) of HCl/Au$_{38}$. An isovalue of 0.01 was used.}
    \label{fig:extra-ks}
\end{figure}

\clearpage

\section*{Relevant Cartesian coordinates}

\begin{verbatim}
Periodic model, HCl/Au(111). Fractional coordinates, Å. CONTCAR file.    
   1.0000000000000000     
     5.8747253417999996    0.0000000000000000    0.0000000000000000
    -2.9373626708999998    5.0876613862999998    0.0000000000000000
     0.0000000000000000    0.0000000000000000   23.1950397491000011
   Au   H    Cl
    16     1     1
Direct
  0.0000000000000000  0.0000000000000000  0.0000000000000000 
  0.4999998857989141  0.0000000000000000  0.0000000000000000   
  0.9999999170361065  0.5000000603125798  0.0000000000000000   
  0.4999999730557505  0.5000000603125798  0.0000000000000000   
  0.1667817367591056  0.3334511060149268  0.1039026481854677   
  0.1666982901287568  0.8333964660564348  0.1040294836894518   
  0.6666694252754581  0.3334511060149268  0.1039026481854677   
  0.6666769793330047  0.8333540728670883  0.1039613623821793   
  0.3334690058975340  0.1669381259961540  0.2064450235746662   
  0.3334442477173170  0.6667137901196512  0.2064639936143564   
  0.8334517243430213  0.1669034508478333  0.2062897869009888   
  0.8332695984219711  0.6667137901196512  0.2064639936143564   
  0.9999835710512741  0.0003697214426523  0.3103389710712889   
  0.5003860361902923  0.0003697214426523  0.3103389710712889   
  0.0000093860876831  0.5000189984157331  0.3102925214402106   
  0.4998100666912819  0.4996202475836355  0.3105421123380978   
  0.6815256679304440  0.8630514500619881  0.4279554222552449 
  0.5947333841464939  0.6894668824940666  0.4685883078379192 
\end{verbatim}

\clearpage

\begin{verbatim}
$comment
Cluster model, HCl/Au10. Cartesian coordinates, Å. Q-Chem input file.
$end

$molecule
!High level atoms
Au     4.40600     5.93700      4.78860
Au     2.93790     8.47970      4.78880
Au     5.87420     8.47970      4.78880
Au     4.40600     7.62960      7.20300
H       4.40610     9.47860      9.92650
Cl      4.40600     8.59550     10.86900
!Low level atoms
Au     7.34340      7.63160      7.19720
Au     5.87350    10.17720      7.19830
Au     2.93860    10.17720      7.19830
Au     5.87590      5.08950      7.19830
Au     1.46870      7.63160      7.19720
Au     2.93620      5.08950      7.19830
$end
\end{verbatim}

\clearpage

\begin{verbatim}
$comment
Cluster model, HCl/Au38. Cartesian coordinates, Å. Q-Chem input file.
$end

$molecule
!High level atoms
Au     4.40600      5.93700      4.78860
Au     2.93790      8.47970      4.78880
Au     5.87420      8.47970      4.78880
Au     4.40600      7.62960      7.20300
H       4.40610      9.47860      9.92650
Cl      4.40600      8.59550     10.86900
!Low level atoms
Au      7.34340      7.63160      7.19720
Au      5.87350    10.17720      7.19830
Au      2.93860    10.17720      7.19830
Au      5.87590      5.08950      7.19830
Au      1.46870      7.63160      7.19720
Au      2.93620      5.08950      7.19830
Au      2.93690      3.39200      4.78880
Au      1.46870      2.54190      7.20300
Au      1.46870      5.93680      4.78490
Au     -0.00050      8.47970      4.78880
Au      0.00120      5.08950      7.19830
Au     -1.46870      7.62960      7.20300
Au     -0.00120    10.17720      7.19830
Au      4.40600      4.24010      2.41300
Au      5.87520      3.39200      4.78880
Au      4.40600      2.54390      7.19720
Au      7.34340      2.54190      7.20300
Au      4.40600      7.63150      0.00000
Au      2.93770      6.78410      2.41000
Au      1.46870      9.32770      2.41300
Au      5.87440      6.78410      2.41000
Au      4.40610      9.32750      2.41140
Au      7.34340      5.93680      4.78490
Au      8.81090      5.08950      7.19830
Au      1.46870    11.02470      4.78860
Au      4.40600    11.02450      4.78490
Au      1.46870    12.71720      7.20300
Au      7.34340      9.32770      2.41300
Au      8.81260      8.47970      4.78880
Au     10.28080     7.62960      7.20300
Au      7.34340    11.02470      4.78860
Au      8.81330    10.17720      7.19830
Au      4.40600    12.71930      7.19720
Au      7.34340    12.71720      7.20300   
$end
\end{verbatim}